\documentclass{pasa}%
\usepackage{graphicx}
\usepackage{amsmath}
\usepackage{amssymb}
\usepackage{lscape}
\usepackage{color}  
\usepackage{natbib}

\def\be{\begin{equation}}
\def\ee{\end{equation}}
\def\bea{\begin{eqnarray}}
\def\eea{\end{eqnarray}}

\def\ms{\noalign{\vskip3pt}}

\def\lapprox{\mathrel{\hbox{\rlap{\hbox{\lower4pt\hbox{$\sim$}}}\hbox{$<$}}}}
\def\gapprox{\mathrel{\hbox{\rlap{\hbox{\lower3pt\hbox{$\sim$}}}\hbox{$>$}}}}

\newcommand{\bphi}{\hbox{\myfont \symbol{30} }}

\newfont{\myfont}{cmmib10}
\newcommand{\bomega}{\hbox{\myfont \symbol{33} }}

\title[Visibility of pulsar emission]{Visibility of pulsar emission: motion of the visible point}
\author[Yuen et al.]{R. Yuen$^{1,2,3,4}$ \and D. B. Melrose$^{1}$\\
\affil{$^1$School of Physics, University of Sydney, Sydney, NSW 2006, Australia}
\affil{$^2$CSIRO Astronomy and Space Science, Australia Telescope National Facility, P.O. Box 76, Epping, NSW 1710, Australia}
\affil{$^3$Xinjiang Astronomical Observatory, Chinese Academy of Science, 40-5 South Beijing Road, Urumqi, Xinjiang, 830011, China}
\affil{$^4$Email: ryuen@xao.ac.cn}}
\jid{PASA}
\doi{xx.xxxx/pas.\the\year.xxx}
\jyear{\the\year}

\bibpunct{(}{)}{;}{a}{}{,}
\begin{document}%
\begin{abstract}
A standard model for the visibility of pulsar radio emission is based on the assumption that the emission is confined to a narrow cone about the tangent to a dipolar field line. The widely accepted rotating vector model (RVM) is an approximation in which the line of sight is fixed and the field line is not strictly tangent to it. We refer to an exact treatment \citep{Gangadhara04} as the tangent model. In the tangent model (but not in the RVM) the visible point changes as a function of pulsar rotational phase, $\psi$, defining a trajectory on a sphere of radius $r$. We solve for the trajectory and for the angular velocity of the visible point around it. We note the recent claim that this motion is observable using interstellar holography \citep{PMBD14}. We estimate the error introduced by use of the RVM and find that it is significant for pulsars with emission over a wide range of $\psi$. The RVM tends to underestimate the range of $\psi$ over which emission is visible. We suggest that the geometry alone strongly favors the visible pulsar radio being emitted at a heights more than ten percent of the light-cylinder distance, where our neglect of retardation effects becomes significant.
\end{abstract}
\maketitle%

\section{INTRODUCTION}
\label{sect:Introduction}

The visibility of pulsar emission is a geometric problem: to identify the ``visible point'' in the pulsar magnetosphere that an observer can see. A standard model for pulsar visibility is based on three assumptions: (i) at the source, the emission is confined to a narrow beam around the direction tangent to the local magnetic field line,  (ii) the magnetic field is dipolar, and (iii) emission occurs only within the open-field region. Let the visible point be described by its spherical polar coordinates, $r,\theta,\phi$ relative to the rotation axis, or $r,\theta_b,\phi_b$ relative to the magnetic axis. Two angles are assumed to be given for a pulsar: the obliquity angle, $\alpha$, between the magnetic axis and the rotation axis, and the viewing angle, $\zeta$, between the line of sight and the rotation axis. The height of the emission point, described by the radial distance $r$, is not well determined, and one considers the location of the visible point on a sphere of radius $r$. The geometric problem is to determine the visible point in terms of $\theta,\phi$ or $\theta_b,\phi_b$ for given $\alpha,\zeta$ as a function of rotational phase, $\psi=\omega_*t$, where $\omega_*$ is the angular speed of rotation.

It is not widely recognized that two different geometric models are used for different purposes. We refer to these as the RVM (rotating vector model) and the tangent model. The RVM was used by \cite{RC69_RVM} to identify the S-shaped swing in the position angle (PA) of the plane of linear polarization, and the RVM continues to be used for this purpose. In the RVM the line of sight is assumed to pass through the center of the star, and the visible point is identified as its intersection with the sphere  of radius $r$. The RVM should be regarded as an approximation to the standard model because the fixed line of sight through the center of the star is not tangent to the field line (with the exception of the special case where the line of sight also intersects the magnetic pole). The tangent model, which was analyzed by \cite{Gangadhara04}, is exact in the sense that the line of sight varies so that it is always tangent to the field line. 

The error introduced by use of the RVM may be understood as follows. For an observer at infinity, $\zeta$ specifies a direction, and hence an infinite set of parallel lines. One of these lines, the one that passes through the center of the star, is chosen in the RVM, and another of these lines, the one that instantaneously satisfies (i), is chosen in the tangent model.  As the pulsar rotates, the line in the RVM remains fixed, implying that the visible point is stationary (fixed $\theta,\phi$). In the tangent model, the line through the visible point changes with $\psi$. The path on the sphere is the trajectory, described by $\theta,\phi$ as a function of $\psi$. Compared with the tangent model, the RVM introduces an angular error. One estimate of this error is the angle between the magnetic field line and the radial vector (the assumed line of sight in the RVM) at the emission point: the RVM is valid only to zeroth order in an expansion in this angle.\footnote{Specifically $\arcsin[\sin\theta_b/(1+3\cos^2\theta_b)^{1/2}]\approx\theta_b/2$ for $\theta_b\ll1$.} Another estimate of the angular error introduced is $\arcsin(d/r)$, where $d$ is the perpendicular displacement between the lines to the observer in the two models. The RVM also introduces a conceptual error: that the visible point is fixed. In the tangent model, the visible point moves, and its motion defines a trajectory on the sphere of radius $r$. 

Our main purpose of this paper is to discuss the implications, for the interpretation of pulsar radio emission, of the existence of this trajectory and the motion of the visible point around it. (Note that neither the trajectory nor the associated motion around it exists in the RVM.) Besides application to conventional observations (duty cycle, interpulse, drifting subpulses, swing of PA), a new possibility is direct measurement of the motion of the visible point. Recent detection of motion with sub-nano-arcsecond accuracy using interstellar holography \citep{PMBD14} shows that this is a realistic possibility. These authors claim that the ``direct observable is the apparent motion of the emission region as a function of pulse phase.'' This (plausibly) corresponds to a direct measurement of the velocity of the visible point, projected along scattering axis. The estimated height of the emission in this particular observation suggests a source well inside the light cylinder radius, $r\ll r_L$, where the magnetic field is well approximated by its dipolar ($\propto1/r^3$) component. 

In Section~\ref{sect:trajectory} we use an analytic solution for the tangent model to derive examples of the trajectory of the visible point in terms of $\theta,\phi$ as functions of $\psi$ for given $\zeta,\alpha$. In Section~\ref{sect:velocity} we discuss the angular velocity of the visible point around the trajectory. In Section~\ref{sect:path} we discuss the significance of assumption (iii), specifically the requirement that the trajectory passes the open-field region. In Section~\ref{sect:swing} we compare the predicted swing of PA in the RVM and the tangent model, and we also comment on the visibility of the opposite pole in the two models. We discuss limitations of the model and summarize our conclusion in Section~\ref{sect:conclusions}.

\section{TRAJECTORY OF THE VISIBLE POINT}
\label{sect:trajectory}

The tangent model has an analytic solution for a strictly dipolar field.

\begin{figure}
\centering  
\includegraphics[width=0.9\columnwidth]{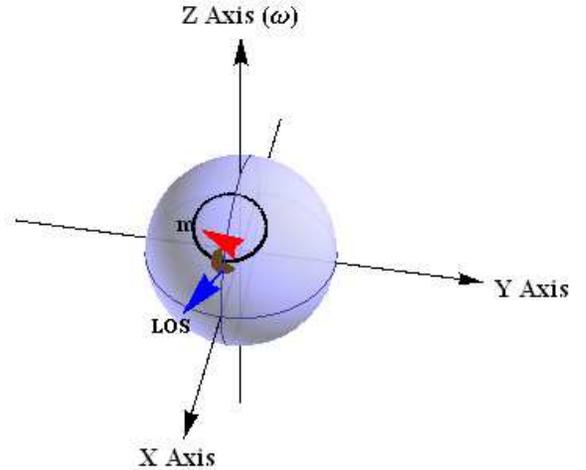}
\caption{This figure shows the viewing geometry of emission in three dimensions for $\alpha=45^\circ$, $\zeta=60^\circ$. A unit sphere, which represents a pulsar magnetosphere, is plotted in Cartesian coordinates. The rotation axis $\bomega$ is along the $z$-axis, the red arrow represents the magnetic moment (${\bf m}$), and the line of sight (LOS) is represented by the blue arrow. The point in brown is the visible point as seen by the observer, in this case, it is at $\psi = 0$ with $\theta_{\rm V} \approx 55^\circ$. The visible point moves as the pulsar rotates from $-\pi$ to $\pi$ tracing out the dark curve.}
\label{fig-TPA45T60}
\end{figure}

\begin{figure}
\centering  
\includegraphics[width=0.9\columnwidth]{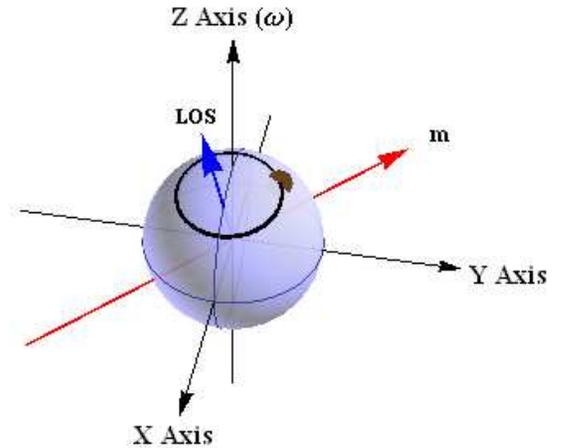}
\caption{As for Figure \ref{fig-TPA45T60}, but for $\alpha=80^\circ$, $\zeta=30^\circ$.}
\label{fig-TPA80T30}
\end{figure}

\subsection{Analytic model for the visible point}

We determine the visible point in terms of $\theta = \theta_{\rm V} (\alpha,\zeta;\psi)$ and $\phi = \phi_{\rm V} (\alpha,\zeta;\psi)$, or $\theta_b = \theta_{b{\rm V}} (\alpha,\zeta;\psi)$ and $\phi_b = \phi_{b{\rm V}} (\alpha,\zeta;\psi)$ by requiring that the line of sight be tangent to a field line,
\be
({\hat{\bf b}}\cdot{\hat{\bf n}})^2=1,
\qquad
{\hat\bphi}_b\cdot{\hat{\bf n}}=0,
\label{dipolar3}
\ee 
where $\hat{\bf b}$ is the unit vector along the magnetic field and $\hat{\bf n} = \sin\zeta\,{\hat{\bf x}}+\cos\zeta\,{\hat{\bf z}}$ is the unit vector along the line of sight. 

Solving Equation (\ref{dipolar3}) simultaneously in the magnetic frame gives \citep{Gangadhara04, Gangadhara05}
\begin{eqnarray}
\cos 2\theta_{b{\rm V}} = \frac{1}{3} (\cos\theta_b \sqrt{8+\cos^2\theta_b} - \sin^2\theta_b), & \nonumber \\
\tan\phi_{b{\rm V}} = \frac{\sin\zeta\sin\psi}{\sin\alpha\cos\zeta-\cos\alpha\sin\zeta\cos\psi}.
\label{EmissionPtPhiB}
\end{eqnarray}
The angles $\theta,\phi$  and $\theta_b,\phi_b$ are related by
\begin{align}
\cos\theta_b&=\cos\alpha\cos\theta+\sin\alpha\sin\theta\cos(\phi-\psi),
\label{transf4}
\\
\ms
\tan\phi_b&=\frac{\sin\theta\sin(\phi-\psi)}{\cos\alpha\sin\theta\cos(\phi-\psi)-\sin\alpha\cos\theta},
\label{transf5}
\end{align}
or
\begin{align}
\cos\theta&=\cos\alpha\cos\theta_b-\sin\alpha\sin\theta_b\cos_b,
\label{transf7}
\\
\ms
\tan(\phi-\psi) &= \frac{\sin\theta_b\sin\phi_b}{\cos\alpha\sin\theta_b\cos\phi_b + \sin\alpha\cos\theta_b}.
\label{transf8}
\end{align}
We choose the zeros of all three azimuthal angles to coincide:  $\phi=\psi=\phi_b=0$. 

\subsection{Examples of the trajectory}

We use ${\it Mathematica} \textsuperscript{\textregistered}$ to plot the visible point for $\psi:=[-\pi,\pi]$ for chosen values of $\alpha:=[0,\pi/2]$ and $\zeta:=[0,\pi/2]$. A point defined by $(\theta,\phi)$ is plotted on the unit sphere corresponding to Cartesian coordinates $\{\sin\theta\cos\phi, \,\sin\theta\sin\phi, \,\cos\theta\}$. 

As $\psi$ varies from $-180^\circ$ to $180^\circ$ the visible point traces a continuous trajectory on the surface of the sphere. Figure \ref{fig-TPA45T60} shows the visible point for $\alpha=45^\circ$ and $\zeta=60^\circ$, with the brown dot at $\psi=0$ and the trajectory shown by the dark closed curve (it is not a circle in general). A further example of the visible point and the trajectory are shown for $\alpha=80^\circ, \zeta=30^\circ$ in Figure \ref{fig-TPA80T30}. A special case is for an orthogonal rotator, $\alpha=90^\circ$, observed along the rotation axis, $\zeta=0^\circ$; the trajectory is then circularly symmetric about the rotation axis, with the line of sight located inside the emission circle for $\alpha>\zeta$.

\section{VELOCITY OF THE VISIBLE POINT}
\label{sect:velocity}

The velocity of the visible point, which is now of direct observational interest \citep{PMBD14}, may be determined by differentiating the solution for the visible point with respect to $\psi$. (This velocity is identically zero in the RVM.) 

\subsection{Definition of the velocity}

The visible point moves at an angular velocity $\bomega_{\rm V}$ with components 
\begin{equation} \label{eq:VelEmPtComp}
\omega_{{\rm V}\theta} = \omega_\star \frac{\partial\theta(\alpha,\psi)}{\partial\psi}, \quad \omega_{{\rm V}\phi} = \omega_\star \frac{\partial\phi(\alpha,\psi)}{\partial\psi}, 
\end{equation}
where $\omega_\star = d\psi/dt$ is the angular speed of the star. The angular speed of the visible point is $\omega_{\rm V} = |\bomega_{\rm V}|$. The motion of the visible point is periodic, with the same period as the star, but the motion can be far from uniform. The motion of the visible point corresponds to sub-rotation around $\psi = 0$, and super-rotation around $\psi=\pi$, with an average $\langle \omega_{\rm V} (\psi) \rangle = \omega_\star$. 

It is instructive to consider the change in velocity as $\alpha$ is decreased; in the aligned case $\alpha=0$ it can be shown that the visible point is stationary. The size of the trajectory decreases, with decreasing $\alpha$, and the asymmetry in the speed of the visible point increases, becoming slower near $\psi=0$ and faster near $\psi=\pi$. In the limit  $\alpha\to0$, the visible point is nearly stationary for nearly all $\psi$, with the exception being very rapid motion near $\psi=\pi$ around a tiny trajectory.

\subsection{Numerical calculation of $\bomega_{\rm V}$}

\begin{figure}
\centering  
\includegraphics[width=1\columnwidth]{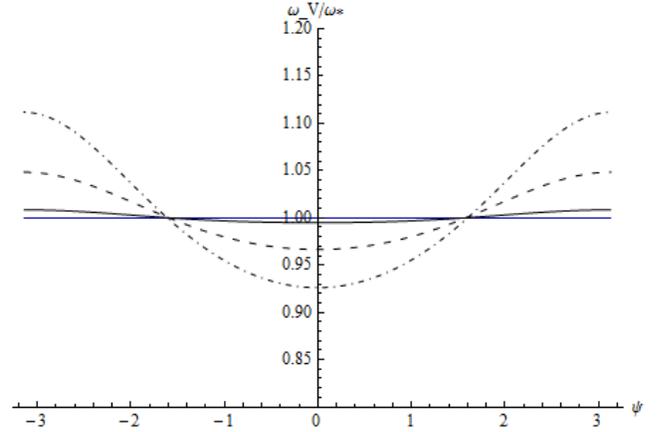}
\caption{The ratio of the angular frequency of the visible point,
$\omega_{\rm V}$, to the spin frequency of the pulsar, $\omega_\star$, plotted
against the rotational phase for $\alpha = 90^\circ, \zeta = 0$ (blue),
$\alpha = 10^\circ, \zeta = 5^\circ$ (solid), $\alpha = 30^\circ, \zeta = 10^\circ$ (dashed), and $\alpha = 45^\circ, \zeta = 15^\circ$ (dot-dashed). The periodic motion over one pulsar period results in $\langle \omega_{\rm V}(\psi) \rangle = \omega_\star$.}
\label{fig-EmPtSpeed_A90A10A30A45}
\end{figure}

Figure \ref{fig-EmPtSpeed_A90A10A30A45} shows the variations of $\omega_{\rm V}$ in units of $\omega_\star$ as a function of rotational phase for various combinations of $\alpha$ and $\zeta$. For $\alpha = 90^\circ$ and $\zeta = 0$, the trajectory is a circle centered at the rotation axis, giving $\omega_{\rm V} = \omega_\star$. For other values of $\zeta$ and $\alpha$, the maximum and minimum values of $\omega_{\rm V}/\omega_\star$ occur at $\psi = 180^\circ$ and 0, respectively. The extrema increase in magnitude with increasing $\alpha$ or $\zeta$, maximizing at $\alpha = \zeta = 90^\circ$. All curves intersect at $\psi \sim \pm 90^\circ$ with $\omega_{\rm V}(\pm90^\circ) / \omega_\star = 1$.

\section{PATH OF THE LINE OF SIGHT}
\label{sect:path}

Assumption (iii) in the tangent model requires that for pulsar emission to be visible the trajectory  must be at least partly inside the open-field region. The trajectory enters the open-field region only if $r$ exceeds a minimum value for a given open field line. The minimum visible height, $r_{\rm min}$, is for the last closed field line.

\subsection{Last closed field line}

When only the dipolar term is retained, the last closed field line is determined by the condition that the field line be tangent to the light cylinder, at $r\sin\theta=r_L$. For given field line, $r=r_0\sin^2\theta_b$, $\phi_b=\phi_0$, this requires
\be
\left.{\partial(\sin^2\theta_b\sin\theta)\over\partial\theta_b}\right|_{\phi_b}=0.
\label{lcfl1}
\ee
The derivative $\partial\theta/\partial\theta_b$ may be determined using (\ref{transf7}), and the resulting equation solved for $\theta_b=\theta_{bL}(\phi_b)$. The value of $\theta\to\theta_L(\phi_b)$ along the last closed field line follows from $\theta_b\to\theta_{bL}(\phi_b)$ in (\ref{transf7}). The shape of the boundary of the open-field region is independent of $r$, and is given by plotting the function $\theta_b=\theta_{bL}(\phi_b)$. The field line constant, $r_0\to r_{L0}(\phi_b)$, for the last closed field line is then determined by
\be
r_{L0}(\phi_b)={r_L\over\sin^2\theta_{bL}(\phi_b)\sin\theta_L(\phi_b)}.
\label{lcf2}
\ee
The radial position along the last closed field line at $\phi_b$ is $r(\theta_b,\phi_b)=r_{L0}(\phi_b)\sin^2\theta_b$.

\subsection{Minimum visible height}

\begin{figure}
\centering  
\includegraphics[width=0.9\columnwidth]{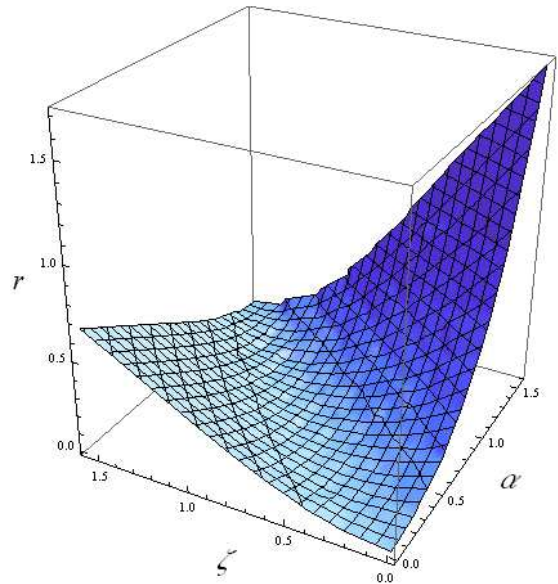}
\caption{A three-dimensional surface plot of $r_{\rm min}$ as functions of $\zeta$ and $\alpha$. The $r_{\rm min}$ increases as $|\beta|$ increases and $r_{\rm min} \rightarrow 0$ for $\beta \rightarrow 0$.}
\label{fig-MinimumViewingAltitudes}
\end{figure}

The minimum visible height, $r_{\rm min}$, is determined by assuming that the trajectory is tangent to the locus of last closed field lines. This corresponds to $\theta_{b{\rm V}}=\theta_ {bL}(\phi_b)$ at $\phi_b=\phi_{b{\rm V}}$, and reproduces the result given by \cite{Gangadhara04}:
\be
r_{\rm min} = \frac{r_L  \sin^2 \theta_{b{\rm V}}}{\sin^2 \theta_{bL}(\phi_{b{\rm V}})\, \sin\theta_L(\phi_{b{\rm V}})}.
\label{MinScrHeight}
\ee
One may regard $r_{\rm min}$ as a function of $\alpha$ and $\zeta$. Figure \ref{fig-MinimumViewingAltitudes} shows a three-dimensional plot for $r_{\rm min}$ as functions of these variables. 

\subsection{Probability of seeing a pulsar}

The existence of a minimum height, below which any emission is not visible (because it is not along the observer's line of sight) has a statistical implication. Consider a populations of neutron stars with $\alpha$ and $\zeta$ randomly distributed. This implies that the impact parameter, $\beta = \zeta - \alpha$, is also randomly distributed. Emission from low heights, $r \ll r_L$, is restricted to a small cone, with half angle $\theta_c$, equal to $(r/r_L)^{1/2}$ in the aligned case, and is visible only for $|\beta|<\theta_c$. The probability of seeing emission from a particular neutron star is given by dividing the solid angle $2\pi\theta_c^2$ (for two poles) by $4\pi$. This probability is $r/2r_L$ for $r/r_L\ll1$. Based on the geometry alone, most visible emission from pulsars must come from relatively large heights. For example, if one assumes that at least 10\% of all radio pulsars are visible, this implies $r/r_L\gtrsim0.2$. This analytic estimate is supported by the numerical results in Figure \ref{fig-MinimumViewingAltitudes}, which show that near the minimum, $r_{\rm min}/r_L$ increases approximately proportional to $\beta^2$, implying that emission from low heights is visible only for small $\beta$.

\subsection{Duty cycle}
\label{sect:ViewGeo}

\begin{figure}
\begin{center}  
\includegraphics[width=0.9 \columnwidth]{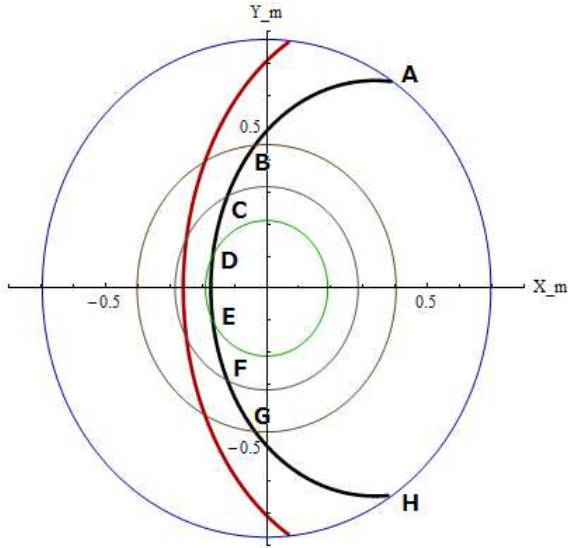}
\caption{The viewing geometry in the magnetic frame showing the observable visible point traces out a path through the open-field region. Four open regions are plotted for $r = 1.2\, r_{\rm min}$ (green), $0.1 r_L$ (gray), $0.2 r_L$ (brown) and $0.6 r_L$ (blue) for $\alpha = 45^\circ$ and $\zeta = 60^\circ$. The trajectory of the visible point (black) intersects the green, gray, brown and blue curves at (D, E), (C, F), (B, G) and (A, H), respectively, between which an observer sees radiation. The red curve represents the path that traces out by the visible point in the conventional model, in which the line of sight is assumed to go through an origin.}
\label{fig-ViewGeo_D}
\end{center}
\end{figure}

In the tangent model, the duty cycle of a pulsar is identified as the range of $\psi$ between the two intersection points where the trajectory cuts the boundary of the open-field region. Figure \ref{fig-ViewGeo_D} shows the viewing geometry in the magnetic frame. Along this portion of the trajectory, $\omega_{\rm V}$ varies symmetrically about $\psi = 0$ (which defines the $x_m$-axis); $\omega_{\rm V}$ is minimum at $\psi = 0$, and increases towards either of the two intersections. The visible point enters the open region, at A, with a certain angular speed, its speed gradually reduces, reaching its minimum at $\psi = 0$, and then increases again until it leaves the open-field region at H with the same angular speed as at A. The portion of the trajectory that is within the open-field region, and hence the duty cycle, increases with increasing $r>r_{\rm min}$.

\section{SWING OF POSITION ANGLE}
\label{sect:swing}

Conventional treatments of the swing of the PA are based on the RVM, and it is important to identify the error that this introduces. Provided this error is small, the RVM is a useful simple approximation to the tangent model. We find that the error is largest near the limits of the duty cycle, and that this affects the predicted visibility of the other pole, interpreted as an interpulse.

\subsection{Evolution of the PA}

\begin{figure}
\begin{center}  
\includegraphics[width=1.0 \columnwidth]{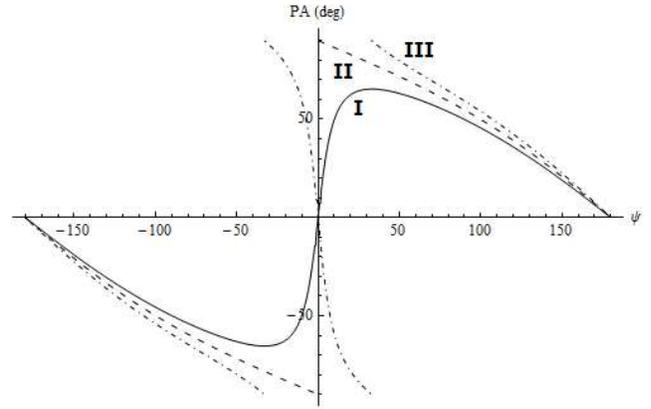}
\caption{The changes in polarization position angle for $\alpha = 45^\circ$ and $\zeta = 40^\circ$ (solid), $\zeta = 45^\circ$ (dashed) and $\zeta = 50^\circ$ (dot-dashed) plotted against rotational phase. Integrated profiles, which are centered at $\psi = 0$ in the model, of different widths capture different information of the PA curve.}
\label{fig-PA_three}
\end{center}
\end{figure}

Pulsar radio emission has a linearly polarized component, described by its PA, which is assumed to be determined by the projection (perpendicular to the line of sight) of the magnetic field line at the visible point. The evolution of the PA with $\psi$ in the RVM is assumed to give a characteristic S-shaped swing as the open-field region sweeps across the line of sight to the center of the star \citep{RC69_RVM}. The actual shape of the PA curve depends on $\alpha$ and $\beta$. Figure \ref{fig-PA_three} shows the sweep in PA as predicted by the tangent model for $\alpha = 45^\circ$ and three different values of $\zeta$. The PA curve changes from I to III as $\beta$ changes from negative to positive, with II corresponding to $\beta = 0$. For $\beta < 0$, the PA swing is an S-shaped curve, similar to a sine curve for $\beta$ large and negative, with the slope of the S steepening as $|\beta|$ decreases towards zero. For $\beta \ge 0$, the PA variation is monotonic, exhibiting a jump by $180^\circ$, which occurs at $\psi = 0$ for $\beta=0$ (II) and at a non-zero $\psi$ for $\beta>0$ (III). 

The observed PA curve depends on the pulse width. For a narrow pulse width the PA change can resemble an unbroken S-shaped curve, as illustrated by curve I in Figure~\ref{fig-PA_three}, examples being PSR B0136+57, B0628-28 and Vela pulsar \citep{LyneManchester1988, BJK+05}, or an abrupt jump, as illustrated by curve II in Figure~\ref{fig-PA_three}, an example being PSR B0355+54 at 1.4 GHz \citep{GL98}. For a larger pulse width the PA change illustrated by curve III in Figure~\ref{fig-PA_three} is similar to that observed from PSR J0738-4042 and PSR J1243-6423 \citep{KJ06}. The interpretation of PA curves suggests that effects other than geometry alone can be significant: `reversed' PA curves indicate that pulsars rotate in the direction opposite to that we assume, and other unusual PA curves may be an indication of non-dipolar field structure, small scale distortions in the open-field region, or interstellar scattering effects \citep{Karastergiou09}.

\subsection{Comparison of RVM and tangent models}

\begin{figure}
\begin{center}  
\includegraphics[width=.95 \columnwidth]{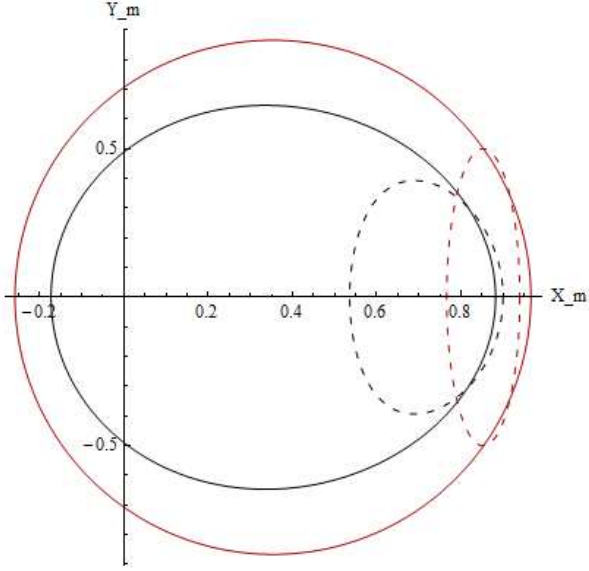}
\caption{The trajectory of the visible point as predicted in  the RVM (red) and the tangent model (black) for $\alpha = 45^\circ$, $\zeta = 60^\circ$ (solid) and $\alpha = 80^\circ$, $\zeta = 30^\circ$ (dashed) in the magnetic frame (see Figures \ref{fig-TPA45T60} and \ref{fig-TPA80T30}). The leftmost point of intersection between a trajectory and the $x_m$ axis represents $\psi = 0$.}
\label{fig-twoTrajs}
\end{center}
\end{figure}

In the RVM the PA swing is  calculated based on the path of the line of sight through the center of the star as the open-field region sweeps across it \citep{RC69_RVM,LyneManchester1988}. An example of the path implied by the RVM is the red curve in Figure \ref{fig-ViewGeo_D}, which is clearly different from the trajectory in the tangent model, shown by the black curve. Further examples showing the difference between the two models are illustrated by the red and black curves, for two different choices of parameters, in Figure \ref{fig-twoTrajs}. Such examples allow one to quantify the error introduced by use of the RVM.

As remarked in Section~\ref{sect:Introduction}, the angular error introduced is $\arcsin(d/r)$, where $d$ is the perpendicular distance between the lines of sight in the RVM and the tangent model. This error is maximum, as a function of $r$, near the minimum value of $r$, at $r=r_{\rm min}$, where it is $d= r_{\rm min} \sin(\theta_b - \theta_{b{\rm V}})$. For given $\zeta$ and $\alpha$,  $d$ has a minimum at $\psi = 0$ and a maximum at $\psi = 180^\circ$, and it increases with increasing $\zeta$ and $\alpha$. 

\begin{figure}
\begin{center}  
\includegraphics[width=.9 \columnwidth]{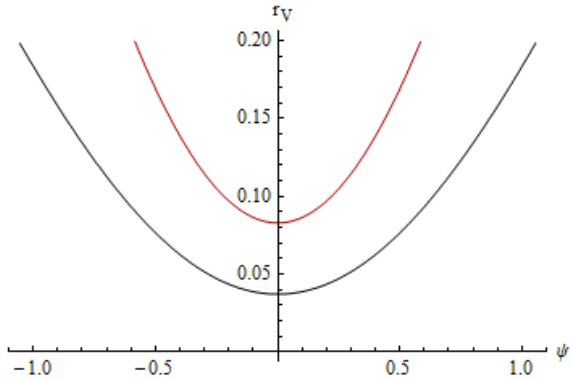}
\caption{Variations in the emission height, $r_V$, along the open field lines where the trajectory cuts through the open region in the RVM (red) and the tangent model (black) for $\alpha = 45^\circ$ and $\zeta = 30^\circ$. The geometry identifies the center of the pulse at $\psi = 0$ where $r_V = r_{\rm min}$, and $r_V = 0.2 r_L$ at the two boundaries representing the leading and trailing edges of the pulse profile.}
\label{fig-VaryEmHeights}
\end{center}
\end{figure}

\begin{figure}
\begin{center}  
\includegraphics[width=1 \columnwidth]{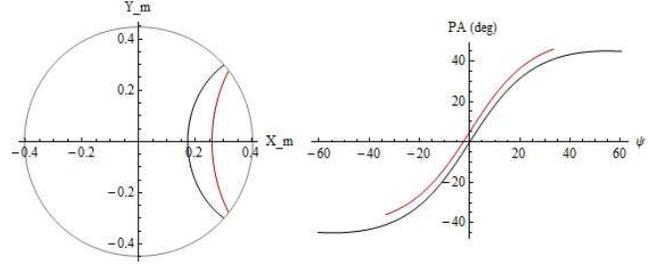}
\caption{Differences in the trajectory through the open-field region in the magnetic frame (left) and the observed PA curve (right) between the RVM (red) and the tangent model (black), for $\alpha = 45^\circ$ and $\zeta = 30^\circ$, and emission height at $0.2 r_L$. The red curve in the PA plot is shifted by $+5^\circ$ for clarity. An S-shaped is predicted for the PA swing in the tangent model.}
\label{fig-TrajPA1}
\end{center}
\end{figure}

\begin{figure}
\begin{center}  
\includegraphics[width=1 \columnwidth]{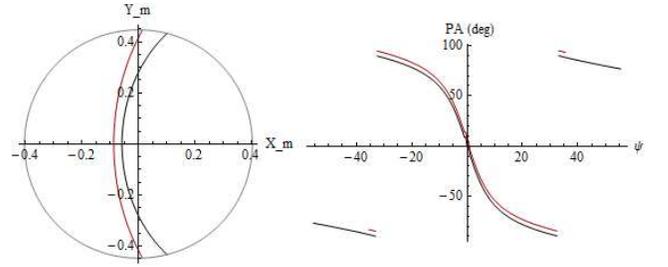}
\caption{Same as in Figure \ref{fig-TrajPA1} but for $\zeta = 50^\circ$.}
\label{fig-TrajPA2}
\end{center}
\end{figure}

\begin{figure}
\begin{center}  
\includegraphics[width=1 \columnwidth]{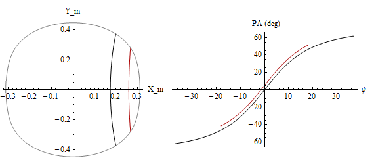}
\caption{Same as in Figure \ref{fig-TrajPA1} but for $\alpha = 80^\circ$ and $\zeta = 65^\circ$. Both with $\beta = -15^\circ$.}
\label{fig-TrajPA3}
\end{center}
\end{figure}

\begin{figure}
\begin{center}  
\includegraphics[width=1 \columnwidth]{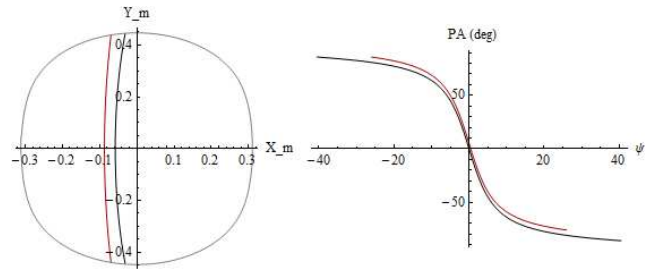}
\caption{Same as in Figure \ref{fig-TrajPA2} but for $\alpha = 80^\circ$ and $\zeta = 85^\circ$. Both with $\beta = 5^\circ$.}
\label{fig-TrajPA4}
\end{center}
\end{figure}

\subsection{Minimum visible height}

Another way of comparing the RVM and the tangent model is in terms of the minimum visible height. The criterion for emission to be visible is that the line of sight, which is independent of $r$ in both models, intersects the open-field region, which broadens with increasing $r$. In both models there is a minimum visible height, at which the line of sight intersects the boundary of the open-field region. Above this height, there is a range of $\psi$ for which emission is visible. In  Figure \ref{fig-VaryEmHeights} we fix this height at a specific value, $r=0.2r_L$, so that this corresponds to the emission height at the two extrema in $\psi$ that define the edges of the pulse window. For $\psi$ between these extrema emission is visible from a range of heights $r_V<0.2r_L$ with this range increasing to a maximum at $\psi=0$, midway between the edges. Figure \ref{fig-VaryEmHeights} compares the pulse window and the range of visible heights within the window for this particular example. The most obvious feature is that the RVM underestimates the size of the pulse window, compared with the tangent model.

\subsection{Swing of PA}

How significant is the error introduced by using the RVM to determine the swing in PA? To answer this question we calculate the swing in PA for both models with given values of $\alpha$ and $\zeta$. The results are shown in Figures \ref{fig-TrajPA1} - \ref{fig-TrajPA4}. The error introduced is small near $\psi=0$, and indeed so small that we need to displace the curves, by $5^\circ$, in order to distinguish between the two models. One can conclude that the RVM is an excellent approximation for the purposes of calculating the PA swing, provided that the pulsar has a narrow pulse window.

The difference in the PA curves between the two models increases as the pulse window increases. The range of $\psi$ predicted by the RVM is narrower than that predicted by the tangent model, and this difference can be substantial for a broad pulse. The largest difference between the two models occurs for large $\beta$. 


In summary, the error introduced by assuming that the line of sight passes through the origin include an underestimation of the height of emission, and hence either to an overestimation of $\alpha$, for $\beta>0$, or to an underestimation of $\alpha$, for $\beta<0$.

\subsection{Visibility of the opposite pole}

The integrated pulse profiles of some pulsars show an interpulse (IP), which is separated from the main pulse by approximately half the rotation period. Similar to main pulses, interpulses may also exhibit such phenomena as mode-changing, pulse-to-pulse intensity modulation and subpulse drifting \citep{Wetal07}. Assuming that the emission mechanism for the near and far magnetic poles is the same and both are actively radiating, the conditions that determine the visibility for the near magnetic pole also apply to visible emission from the far (opposite) pole. The analytic solution in Section~\ref{sect:trajectory} involves solving a quartic equation, and there are four solutions. Two of the solutions correspond to emission in the observer's hemisphere, and the condition for an interpulse is that both be visible. (The other two solutions correspond to emission in the opposite hemisphere to the observer and are of no interest.)

\begin{figure}
\begin{center}  
\includegraphics[width=1 \columnwidth]{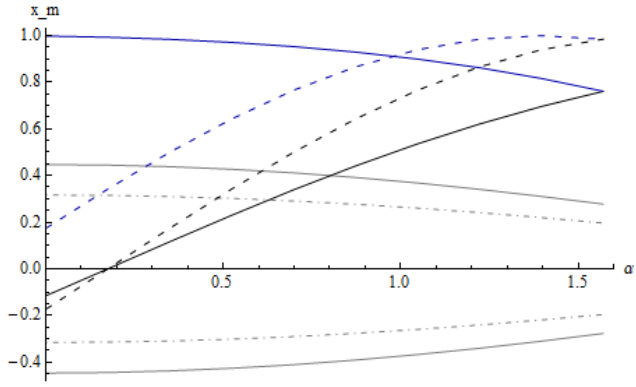}
\caption{The minimum visible height for $\psi = 0$ is shown by the value of $x_m$ for the main pulse (black) and interpulse (blue) in the tangent model (solid) and the RVM (dashed) as a function of $0\le\alpha\le90^\circ$ and $\zeta = 10^\circ$. The boundaries for the open regions are shown for $r = 0.1 r_L$ (dot-dashed gray) and $0.2 r_L$ (solid gray) where each intersects the $x_m$ axis at two points ($x_m>0$ and $x_m<0$, see Figure \ref{fig-TrajPA1}). The two magnetic poles, which are separated by $\theta = \psi = \pi$, are plotted at the same origin for clarity and both are assumed with similar constrains on the emission height, and hence for $\beta = 0$ (intersection of the black curve with the $x_m$-axis), $r_{\rm min} = 0$ for near pole but $r_{\rm min,IP}$ is large. Visible emission requires the visible point to be between the gray lines. Intersection of a curve with the horizontal axis occurs when $\zeta = \alpha= 10^\circ$, but only for the main pulse.
}
\label{fig-Vis_MPIP_Z10}
\end{center}
\end{figure}

\begin{figure}
\begin{center}  
\includegraphics[width=1 \columnwidth]{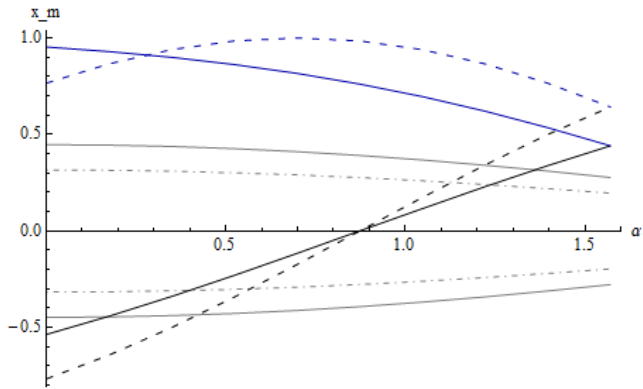}
\caption{Same as in Figure \ref{fig-Vis_MPIP_Z10} but for $\zeta = 50^\circ$; only the near pole is visible.}
\label{fig-Vis_MPIP_Z50}
\end{center}
\end{figure}

\begin{figure}
\begin{center}  
\includegraphics[width=1 \columnwidth]{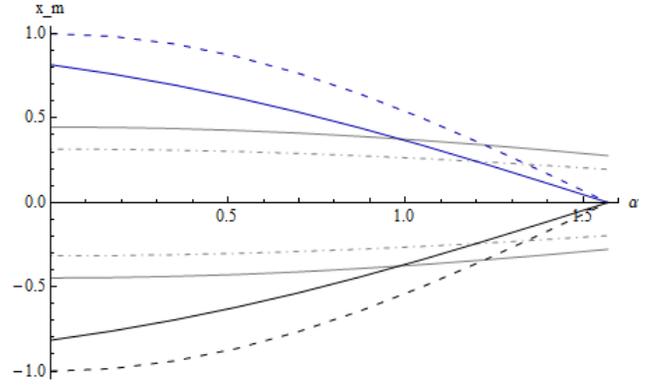}
\caption{Same as in Figure \ref{fig-Vis_MPIP_Z10} but for $\zeta = 90^\circ$; both poles are visible for large $\alpha$. The  RVM (dashed curves) underestimates the range of $\alpha$ for which both poles are visible.}
\label{fig-Vis_MPIP_Z90}
\end{center}
\end{figure}

The visibility conditions for an interpulse are that a portion of the trajectory of the visible point near the far pole is within the open field region, and that this occurs above the minimum visible height, $r_{\rm min,IP}$. Figures \ref{fig-Vis_MPIP_Z10} -- \ref{fig-Vis_MPIP_Z90} apply in the special case $\psi=0$, when the magnetic axis, the rotation axis and the line of sight are in the same plane. These figures show the variations in $r_{\rm min}$ and $r_{\rm min,IP}$, where IP refers to interpulse, as predicted in both the RVM and the tangent model as function of $\alpha$ for three values of $\zeta$. Two emission heights are selected, at $r = 0.1 r_L$ and $r = 0.2 r_L$, with the range of $\pm x_m$ from which emission is visible restricted to between the curves, which are symmetric about the horizontal axis. 

Three features are apparent from Figures \ref{fig-Vis_MPIP_Z10} -- \ref{fig-Vis_MPIP_Z90}. First, the far pole is visible only for large $\alpha$ and $\zeta$, and large heights. Second, from Figure~\ref{fig-Vis_MPIP_Z90} one infers that the RVM underestimates the range of $90^\circ-\alpha$ over which the far pole is visible.  Third, from Figure~\ref{fig-Vis_MPIP_Z10} it is evident that the RVM incorrectly predicts a range of small $\alpha$ and $\zeta$ where the far pole is visible. These differences illustrate the errors that can be introduced by using the RVM  away from $\psi\approx0$, where the error is small.

\begin{figure}
\centering  
\includegraphics[width=0.5\columnwidth]{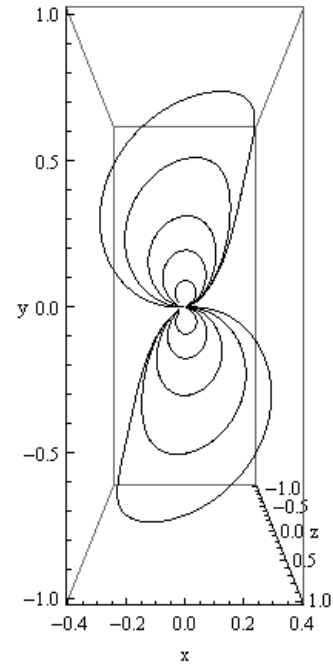}
\caption{Simulation of the magnetic field lines at various $r$ for $\alpha = 90^\circ$ when looking down from the rotation axis showing the dipolar structure for $r \leq 0.2 r_L$.}
\label{fig-FLsA90_Ind2}
\end{figure}

\section{CONCLUSIONS} 
\label{sect:conclusions}

The tangent model for the visibility of pulsar radio emission \citep{Gangadhara04} is widely accepted, but its implications are not widely recognized. The tangent model is inconsistent with the earlier RVM, which continues to be used, especially in connection with the swing of PA. In this paper, we use an analytic solution for the tangent model to demonstrate some of its implications, and to compare the predictions based on it with predictions based on the RVM. Before summarizing our results in dot-point form, we comment on limitations on the tangent model in the form assumed here.

The tangent model we explore neglects many physical effects, with the most important being retardation and aberration  \citep{GG03,Gangadhara05}. The exact solution for a rotating magnetic dipole includes the dipolar term, $\propto1/r^3$, which is the only term we retain, and the inductive ($\propto1/r^2$) and radiative ($\propto1/r$) terms (collectively the retarded terms) that we ignore. Inclusion of the retarded terms leads to well-known distortions in the field \citep{AE98, DH04b}, compared with the dipolar term. The distortions are a strong function of $\alpha$ (being absent for $\alpha=0$), and we illustrate their magnitude by plotting, in Figure~\ref{fig-FLsA90_Ind2}, some of the exact field lines for the special case of $\alpha=90^\circ$. Based on our calculations, and on the work of others,  e.g., \citet{HH97}, we estimate that these distortions are relatively unimportant for  $r \leq 0.2 r_L$. An exact analytic solution for the trajectory is not feasible when the retarded terms are included. It is possible to complement the exact solution for a dipolar field by a perturbation approach to include the effects of retardation. For example, \cite{BCW91} included retardation effects as a perturbations to the RVM. The perturbations should be applied to the tangent model, and not to the RVM, but we do not do so here. Similarly, other field line distortions, the finite size of the emission cone and aberration may be included as perturbation corrections to the exact solution. Again, such perturbations should be applied to the exact solution, and not to the RVM.

\subsection*{Summary}
\begin{itemize}

\item The older model for the visibility of pulsar radio emission \citep{RC69_RVM,LyneManchester1988} is incompatible with the widely accepted model \citep{Gangadhara04} in which radiation is beamed along the magnetic field line at the point of emission. We refer to the two models as the RVM and the tangent model, respectively.

\item The tangent model (but not the RVM) implies that the visible point moves around a trajectory as the pulsar rotates. The size of the trajectory and the speed of the motion of the visible point around it depend on $\alpha$ and $\zeta$.

\item The requirement that emission come only from open field lines sets a minimum height for visible emission, and this is different in the two models.

\item The geometry alone strongly favors visible radio emission being originating at a heights more than ten percent of the light-cylinder distance.

\item The size of the pulse window is different in the two models, with the RVM typically underestimating the range of $\psi$ for which emission is visible.

\item The errors introduced by using the RVM are particularly notable when considering the visibility of the opposite pole, and hence of an interpulse.

\item The swing of PA is similar in the two models provided the pulse window is narrow, but the RVM becomes increasingly unreliable as the size of the pulse window increases.

\end{itemize}

\section*{Acknowledgments} 
We thank Dick Manchester for fruitful discussion, and Patrick Weltevrede for helpful advice.

\end{document}